# Epitaxial lift-off of $La_{2/3}Sr_{1/3}MnO_3$ membranes enabled by BaO sacrificial layers and restoration of the Curie temperature

Takahito Takeda,[1,a)] Daigo Matsubara,[2] Yuki K. Wakabayashi,[3] Kohei Yamagami,[4] Munetoshi Seki,[2,5] Hitoshi Tabata,[2,5] Le Duc Anh,[2,5] Masaki Kobayashi,[3] Masaaki Tanaka,[2,5,6] and Shinobu Ohya[2,5,6,b)]

[1]Department of Chemical System Engineering, The University of Tokyo, 7-3-1 Hongo, Bunkyo-ku, Tokyo 113-8656, Japan

[2]Department of Electrical Engineering and Information Systems, The University of Tokyo, 7-3-1 Hongo, Bunkyo-ku, Tokyo 113-8656, Japan

[3] Basic Research Laboratories, NTT, Inc., Atsugi, Kanagawa 243-0198, Japan

[4]Japan Synchrotron Radiation Research Institute (JASRI), 1-1-1 Kouto, Sayo, Hyogo, 679-5198, Japan

[5]Center for Spintronics Research Network, The University of Tokyo, 7-3-1 Hongo, Bunkyo-ku, Tokyo 113-8656, Japan

[6]Institute for Nano Quantum Information Electronics (NanoQuine), The University of Tokyo, 4-6-1 Komaba, Meguro-ku, Tokyo 153-8505, Japan

a) Author to whom correspondence should be addressed: ttakeda@g.ecc.u-tokyo.ac.jp
b) Author to whom correspondence should be addressed: ohya@cryst.t.u-tokyo.ac.jp

**Abstract**

Ultrathin complex-oxide membranes provide a powerful platform for strain engineering, interfacial control, and heterogeneous integration; however, their formation remains constrained by the availability and performance of suitable water-soluble sacrificial layers. This letter demonstrates that barium oxide (BaO) serves as a highly efficient and rapidly dissolving water-soluble sacrificial layer, enabling the epitaxial lift-off and transfer of ultrathin $La_{2/3}Sr_{1/3}MnO_3$ (LSMO) membranes onto $SiO_x$/Si substrates. LSMO membranes with a thickness of approximately 8 nm are released using a BaO sacrificial layer grown by molecular beam epitaxy, while high crystallinity is preserved and Ba interdiffusion is limited to a narrow interfacial region of approximately 0.5 nm. Post-transfer oxygen annealing at 600 °C increases the Curie temperature ($T_C$) from 342 K to 346 K by eliminating $Mn^{2+}$ states associated with oxygen vacancies generated through oxygen extraction into the BaO layer. These results show that BaO provides a fast, scalable, and compositionally simple route for complex-oxide membrane release, while brief oxygen annealing is essential to restore the optimal Mn valence state and achieve the intrinsic high $T_C$.

Perovskite oxides are key materials for next-generation electronics owing to the strong coupling among their spin, charge, and lattice degrees of freedom.[1–4] Among these materials, La$_{1-x}$Sr$_x$MnO$_3$ stands out for spintronic applications due to its half-metallic character and Curie temperature ($T_C$) exceeding 300 K,[5,6] which originates from double-exchange interactions through the Mn$^{3+}$–O–Mn$^{4+}$ bonding pathway.[7] La$_{2/3}$Sr$_{1/3}$MnO$_3$ (LSMO) has demonstrated exceptional spintronic performance, including tunnel magnetoresistance ratios approaching approximately 1900% in magnetic tunnel junctions[8,9] and giant spin-valve ratios of up to 140% in lateral spin-MOSFETs.[10] However, the performance of LSMO thin films is often limited by interfacial dead layers, where magnetization and electrical conductivity are severely degraded.[11,12] These degraded regions arise from substrate-imposed epitaxial strain and oxygen octahedral rotations.[13–15] Therefore, epitaxial lift-off using water-soluble sacrificial layers, such as Sr$_3$Al$_2$O$_6$ and Sr$_4$Al$_2$O$_7$, has emerged as an effective approach to physically decouple oxide films from their growth substrates. This release relaxes epitaxial constraints, suppresses oxygen octahedral rotations, and enhances the $T_C$ of LSMO by reducing the dead-layer thickness.[16–21] Beyond mitigating interfacial degradation, oxide membranes provide unprecedented post-growth tunability. These membranes can sustain extreme strain states and enable deterministic control through structural engineering, such as process-induced rippling or the stacking and twisting of membrane layers.[22,23] Such capabilities are critical for integrating functional complex oxides, such as LSMO, with Si-based platforms and for advancing flexible, high-performance spintronic technologies.

In this letter, BaO is used as a water-soluble sacrificial layer to prepare LSMO membranes. Compared to established sacrificial materials such as Sr$_3$Al$_2$O$_6$ and Sr$_4$Al$_2$O$_7$, BaO is easier to grow because of its simpler composition, can be grown epitaxially on SrTiO$_3$ (STO), and dissolves more rapidly in water.[24,25] From the perspective of Madelung energy, BaO is expected to exhibit higher water solubility than Sr$_3$Al$_2$O$_6$ and Sr$_4$Al$_2$O$_7$, as BaO consists of divalent Ba$^{2+}$ ions with a larger ionic radius, whereas Sr$_3$Al$_2$O$_6$ and Sr$_4$Al$_2$O$_7$ contain highly charged trivalent Al$^{3+}$ ions with smaller ionic radii, resulting in a highly robust crystal lattice. Using BaO, epitaxial lift-off and transfer of ultrathin LSMO membranes were achieved. The $T_C$ increased from 342 K in the as-transferred state to 346 K after annealing, corresponding to an improvement of 4 K achieved through oxygen annealing. This enhancement originates from the oxidation of Mn$^{2+}$ ions, which are formed by oxygen vacancies generated through oxygen extraction from LSMO into the BaO layer. Post-transfer annealing effectively restores the optimal Mn$^{3+}$/Mn$^{4+}$ ratio, which is essential for robust double-exchange ferromagnetism.

A heterostructure composed of LSMO [20 unit cells (u.c.), approximately 8 nm]/BaO (approximately 18 u.c., approximately 9 nm) was grown on an STO (001) substrate by molecular beam epitaxy (MBE) [Fig. 1(a)]. BaO, with a lattice constant $a = 5.539$ Å, has a pseudocubic lattice constant of $5.539/\sqrt{2} = 3.917$ Å and therefore grows epitaxially on STO ($a = 3.905$ Å) with a 45° in-plane rotation. The lattice mismatch between BaO and STO is -0.3%. Prior to growth, the STO substrate was cleaned with hydrofluoric acid and annealed at 1000 °C to obtain a TiO$_2$-terminated surface. BaO was grown at 540 °C without supplying oxygen or ozone to avoid the formation of BaO$_2$.[26] Reflection high-energy electron diffraction (RHEED) oscillations indicate a layer-by-layer growth mode [Fig. 1(b)], and the streaky RHEED pattern confirms the formation of an atomically flat surface [Fig. 1(c)]. LSMO was grown at 720 °C under a mixed oxygen and ozone atmosphere (80% and 20%, respectively) at a total pressure of 2×10$^{-4}$ Pa using a shuttered growth sequence. The RHEED pattern after LSMO growth also indicates a flat surface [Fig. 1(c)]. Because BaO is highly air sensitive and degrades rapidly under ambient conditions, the crystallinity of the BaO layer was evaluated by growing a 6 nm thick BaO film under identical conditions and capping it with a metal layer for X-ray diffraction (XRD) measurements. Clear fringes in the XRD pattern [Fig. 1(d)] confirm the formation of a high-quality BaO layer. The out-of-plane lattice constant extracted from XRD is 5.54 Å, which is nearly identical to the bulk value of 5.539 Å, consistent with the small lattice mismatch.

The LSMO membranes were obtained from the as-grown LSMO/BaO/STO structure using an epitaxial lift-off process, as illustrated in Fig. 2. A thermal release tape (TRT) was laminated onto the LSMO surface. The sample was immersed in pure water for 1 hour to dissolve the sacrificial BaO layer, releasing the LSMO film onto the TRT. This dissolution process is significantly faster than that of $Sr_3Al_2O_6$ of comparable thickness, which typically requires more than 10 hours.[17,27] This fast-dissolving behavior is advantageous for high-throughput membrane fabrication and for minimizing time-dependent damage during immersion in pure water.[28,29] The TRT/LSMO stack was attached to a $SiO_x$/Si substrate and heated at 80 °C for 10 minutes to raise bonding between LSMO and $SiO_x$. Finally, the TRT was removed by heating to 130 °C, yielding the LSMO membrane transferred onto the $SiO_x$/Si substrate.

Structural and elemental analyses were performed on the as-transferred LSMO membrane using high-angle annular dark-field scanning transmission electron microscopy (HAADF-STEM) and energy-dispersive X-ray spectroscopy (EDX). The as-transferred membrane formed flakes with lateral sizes of up to approximately 100 μm [Fig. 3(a)], and the HAADF-STEM image of a single flake shown in Fig. 3(b) confirms that the transferred LSMO retains high crystallinity. The EDX elemental maps of La, Sr, Mn, and Ba reveal slight interfacial diffusion [Fig. 3(c)], and line profiles along the film thickness show Ba diffusion from the sacrificial layer into the LSMO layer near the LSMO/BaO interface [Fig. 3(d)]. A normalized comparison of the EDX distributions [Fig. 3(e)] further indicates that, while La and Mn exhibit similar profiles, Sr displays a distinct distribution. From the widths of the La (approximately 8.6 nm) and Sr (approximately 8.1 nm) profiles, it is estimated that Ba partially substitutes for Sr sites within approximately 0.5 nm, corresponding to approximately 1 u.c., near the interface. Because $Ba^{2+}$ and $Sr^{2+}$ have the same valence in perovskite oxides, this substitution is expected to have a negligible effect on the Mn valence.

The samples were annealed at 600 °C under 1 atm of pure oxygen for 2 hours to investigate the effect of annealing on the magnetic properties of LSMO membranes. Superconducting quantum interference device measurements show that $T_C$ increases from 342 K in the as-transferred state to 346 K after annealing, whereas the magnetic hysteresis curves remain essentially unchanged by the annealing process [Figs. 4(a) and 4(b)]. The $T_C$ values of the as-transferred and annealed membranes are comparable to those obtained using $Sr_3Al_2O_6$ and $Sr_4Al_2O_7$ sacrificial layers,[16–21] although the $T_C$ of $La_{2/3}Ba_{1/3}MnO_3$ is approximately 20 K lower than that of LSMO.[30–32] This observation indicates that the Ba diffusion observed in this study has only a minor influence on the magnetic properties, likely because the Ba diffusion is confined to the interfacial dead layer formed in the LSMO membranes.[16,18] It is assumed that the Ba distribution is not altered by oxygen annealing, since the annealing temperature of 600 °C is lower than the LSMO growth temperature of 720 °C.

Enhancement of $T_C$ in LSMO thin films by annealing has been previously reported and is attributed to the compensation of oxygen vacancies.[33–35] X-ray absorption spectroscopy (XAS) measurements were performed at SPring-8 BL25SU to clarify the origin of the ferromagnetic enhancement induced by annealing in the LSMO membranes. The Mn $L_{2,3}$ XAS results demonstrate that $Mn^{2+}$ states present in the as-transferred membrane are eliminated after annealing [Fig. 4(c)]. This behavior differs from that observed with $Sr_4Al_2O_7$, where $Mn^{2+}$ is negligible without annealing.[36] This difference is attributed to the BaO growth conditions, specifically the absence of $O_2$ or $O_3$ flux during BaO growth, which likely induces oxygen extraction from the LSMO layer into the BaO layer, generating oxygen vacancies and $Mn^{2+}$ in LSMO. Therefore, annealing in an oxygen atmosphere is indispensable for restoring the optimal Mn valence state and achieving high $T_C$ values when BaO is used as a sacrificial layer for oxide membrane preparation. The fabrication of large-area LSMO membranes free of cracks can be facilitated by strategies such as increasing the LSMO thickness and or introducing a supporting layer that does not compromise the intrinsic properties of LSMO.[37,38]

Accordingly, BaO was employed as a fast-dissolving sacrificial layer for the preparation of LSMO membranes. The transferred LSMO retains high crystallinity with slight Ba diffusion, approximately 1 u.c., into Sr sites at the interface. Post-transfer annealing at 600 °C in an oxygen atmosphere increases $T_C$ from 342 K to 346 K. XAS measurements confirm that this enhancement originates from the elimination of $Mn^{2+}$ states formed by oxygen extraction from LSMO into the BaO layer. These results demonstrate that BaO is an effective sacrificial layer for high-throughput membrane fabrication, provided that oxygen annealing is applied to restore the optimal Mn valence state and achieve high $T_C$.

## ACKNOWLEDGMENTS

This work was supported partly by Grants-in-Aid for Scientific Research (Nos. 25H00840, 24K17303, 24KK0107, 23K19112, 23H03802, 22H04948), FOREST (No. JPMJFR2444), ERATO (JPMJER2202) of the Japan Science and Technology Agency, Spintronics Research Network of Japan (Spin-RNJ), and Iketani Science and Technology Foundation, Japan. Supporting experiments at SPring-8 were approved by the Japan Synchrotron Radiation Research Institute (JASRI) Proposal Review Committee (Proposal Nos. 2024B2441, 2025A1361).

## AUTHOR DECLARATIONS

### Conflict of interest

The authors have no conflicts to disclose.

### Author contributions

**Takahito Takeda**: Conceptualization (lead) ; Funding acquisition (supporting); Data curation (lead); Formal analysis (lead); Investigation (lead); Methodology (lead); Project administration (equal); Visualization (lead); Writing - original draft (lead); Writing - review & editing (lead). **Daigo Matsubara**: Methodology (supporting). **Yuki K. Wakabayashi**: Methodology (supporting). **Kohei Yamagami**: Resources (supporting). **Munetoshi Seki**: Resources (supporting). **Hitoshi Tabata**: Resources (supporting). **Le Duc Anh**: Conceptualization (equal). **Masaki Kobayashi**: Resources (supporting). **Masaaki Tanaka**: Resources (supporting); Funding acquisition (supporting). **Shinobu Ohya**: Conceptualization (equal); Funding acquisition (lead); Project administration (lead); Resources (supporting); Writing - review & editing (equal).

## DATA AVAILABILITY

The data supporting the findings of this study are available from the corresponding author upon reasonable request.

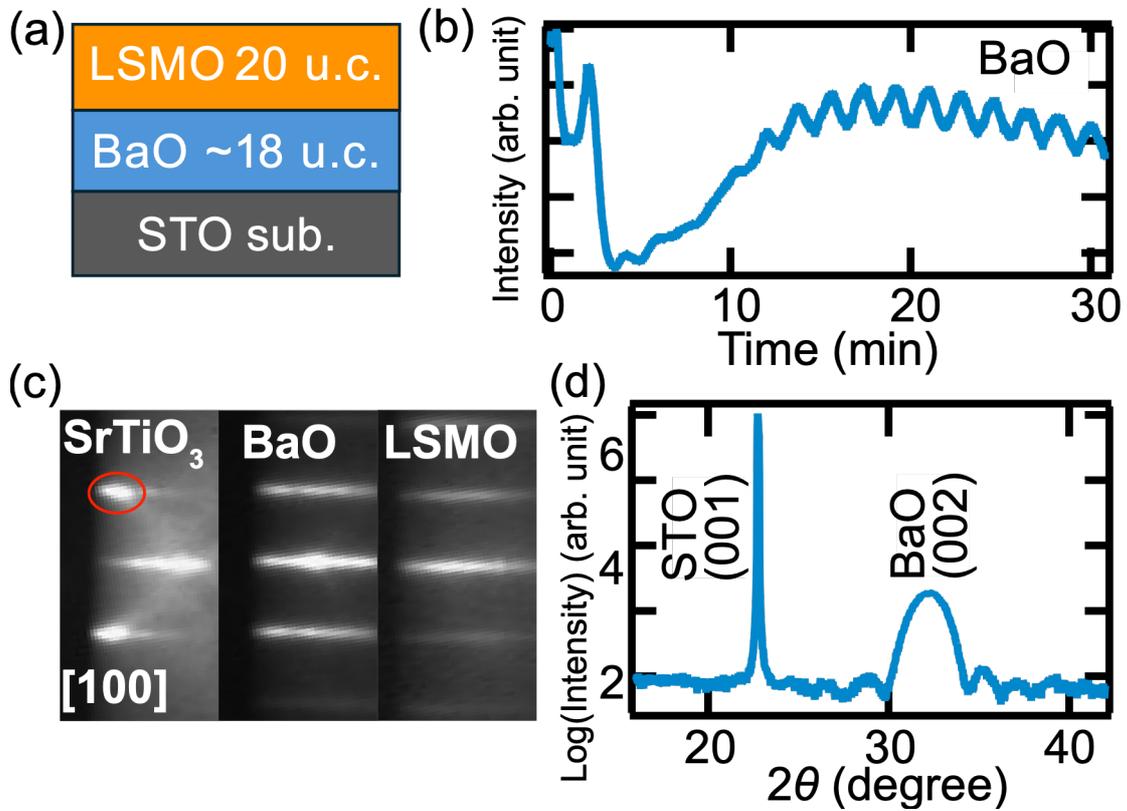

FIG. 1. (a) Schematic cross-sectional structure of the LSMO/BaO/STO sample. (b) RHEED oscillations obtained by monitoring the (10) spot (circled by red in (c)) during BaO growth by MBE. (c) RHEED patterns of the STO substrate, BaO (18 u.c.), and LSMO (20 u.c.) taken along the [100] direction of the STO substrate. (d) XRD pattern of the metal-capped BaO (6 nm)/STO sample.

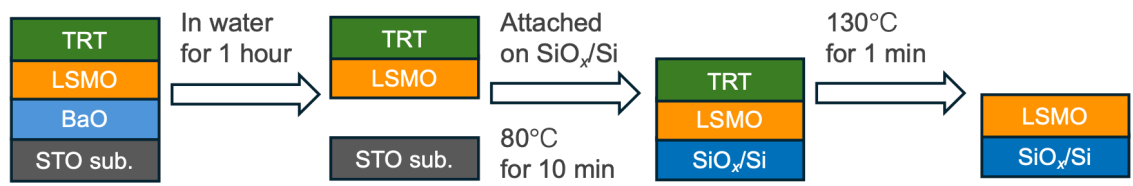

FIG. 2. Schematic illustration of the LSMO transferring process.

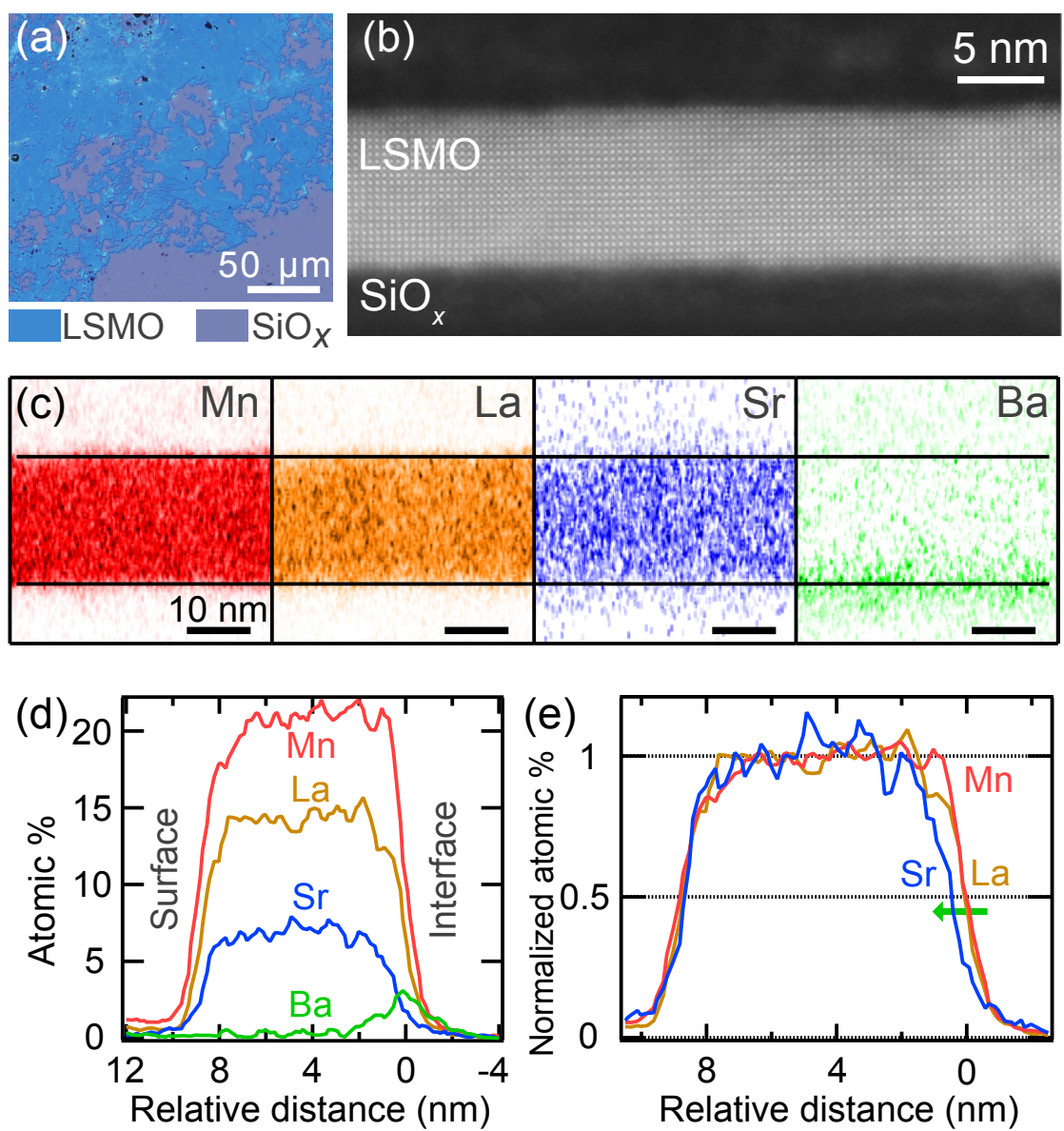

FIG. 3. (a) Optical microscope image of the LSMO membranes on a $SiO_x$/Si substrate. (b) HAADF-STEM image of an LSMO membrane with the electron beam azimuth along the [100] direction of LSMO. (c) EDX elemental maps of Mn (red), La (orange), Sr (blue), and Ba (green). (d) Depth profiles of the EDX intensities of Mn (red), La (orange), Sr (blue), and Ba (green). (e) Normalized depth profiles of Mn, La, and Sr corresponding to Fig. 3(d).

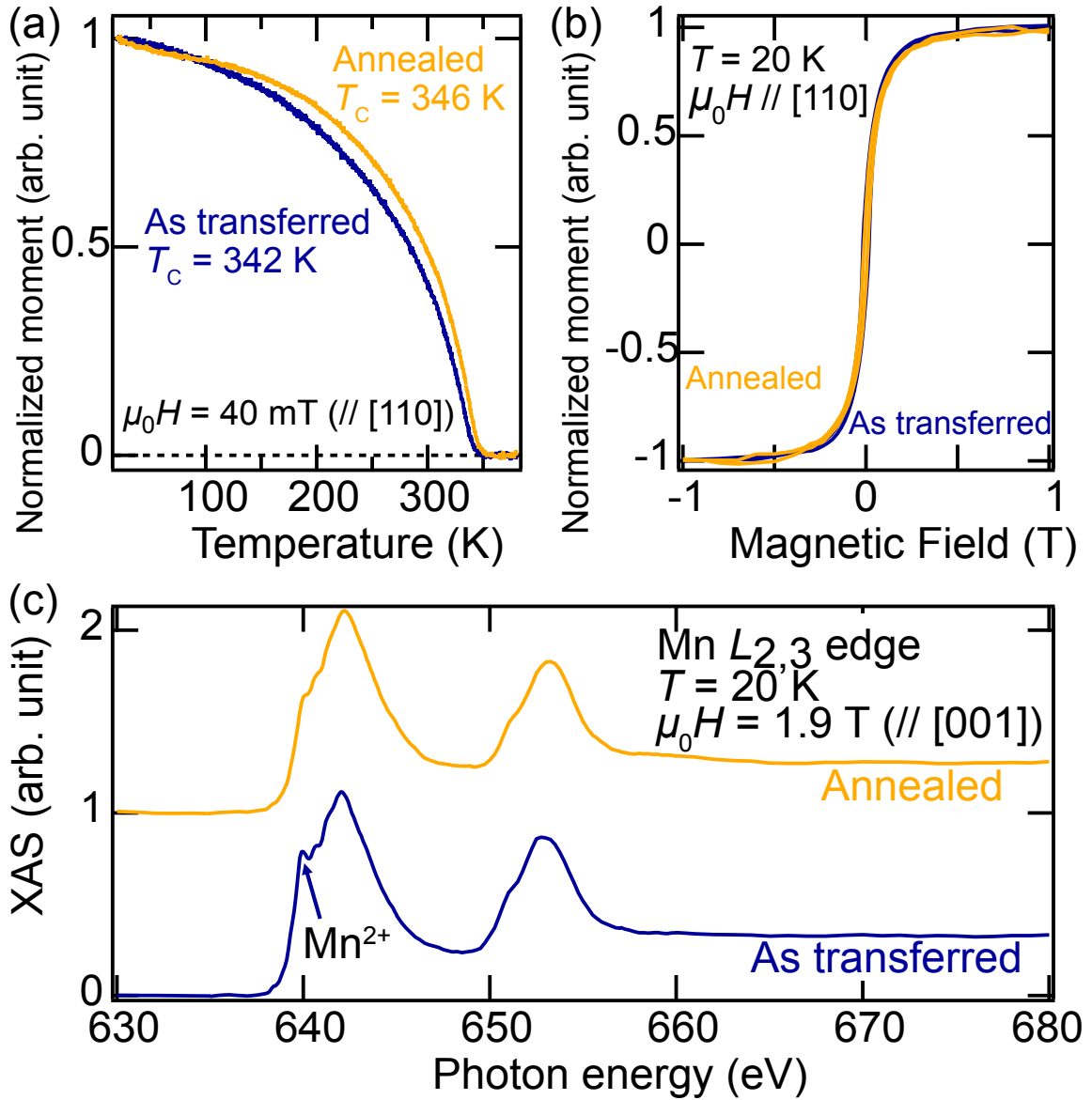

FIG. 4. (a) Temperature dependence of the normalized magnetization for the as-transferred and annealed LSMO membranes, measured in a magnetic field of 40 mT applied along the in-plane [110] direction after field cooling of 1 T. (b) Normalized magnetization at 20 K for the as-transferred and annealed LSMO membranes as a function of a magnetic field applied along the in-plane [110] direction. (c) XAS spectra at the Mn $L_{2,3}$ absorption edges measured at 20 K in a magnetic field of 1.9 T applied perpendicular to the film plane (// [001]).